\documentclass[aps,prl,twocolumn,showpacs,superscriptaddress]{revtex4}
\usepackage{graphicx,color}

\usepackage{url}
\usepackage{helvet}

\begin{document}

\title{Multi-Cascade Proton Acceleration by Superintense Laser Pulse in the Regime of Relativistically Induced Slab Transparency}

\author{A.A.~Gonoskov}
\author{A.V.~Korzhimanov}
\author{V.I.~Eremin}
\author{A.V.~Kim}
\author{A.M.~Sergeev}
\affiliation{Institute of Applied Physics, Russian Academy of Sciences, 603950 Nizhny Novgorod, Russia}

\date{\today}

\begin{abstract}

A regime of multi-cascade proton acceleration in the interaction of $10^{21}-10^{22}$ W/cm$^2$ laser pulse with a structured target is proposed. The regime is based on the electron charge displacement under the action of laser ponderomotive force and on the effect of relativistically induced slab transparency which allows to realize idea of multi-cascade acceleration. It is shown that a target comprising several properly spaced apart thin foils can optimize the acceleration process and give at the output quasi-monoenergetic beams of protons with energies up to hundreds of MeV with energy spread of just few percent.

\end{abstract}


\maketitle

{\em Introduction.}{\textemdash}Modern laser technology offers a wide range of possibilities for exploring those regimes of interaction between a laser pulse and a plasma where effects such as relativistic and striction nonlinearities play a fundamental role in determining the dynamical evolution of the system (see, for example, \cite{mourou} and references therein). These effects become very important in the proposed scientific and technical applications, which range from plasma-based particle accelerators \cite{tajima} to inertial confinement fusion \cite{tabak,wilks1}, including compact ion accelerators for hadronotherapy in biomedicine and proton imaging \cite{bulanov,borghesi}. The present paper is concerned with the problem of obtaining high-energy proton or light ion monoenergetic beams from the interaction of superintense laser radiation with solid-state targets. It should be noted that this problem is an area of active research during the past decade. The emitted ion and, in particular, proton pulses contain large particle numbers between $10^{10}$ and $10^{13}$ with energies in the multi-MeV range \cite{schwoerer, hegelich, hatchett, mackinnon, willingale, robson, yogo}. Different mechanisms for ion acceleration have been examined in kinetic simulations; these include target normal sheath acceleration (TNSA) at the rear side \cite{wilks2, fuchs}, hole boring and shock acceleration at the front side \cite{pukhov, zhidkov, denavit, silva}. Last year there were published a few papers where new mecanisms of ion acceleration for $10^{21}$-$10^{22}$ W/cm$^2$ laser pulses were actively discussed \cite{yin, robinson, yan, korzhimanov1, zhang}. In opposite to TNSA regime the proposed techniques are based on electron charge displacement due to ponderomotive force of laser radiation which is much more effective when circularly polarized light is used for ion acceleration at the considering intensities.

{\em Concept.}{\textemdash}In the present paper we propose a new regime of ion acceleration that is effective for the laser intensity range of $10^{21}$-$10^{22}$ W/cm$^2$, which is accessible in several labs today. This regime is capable for accelerating both light ions and protons but for short further we will talk only about protons.

The proposed regime consists of three phases of acceleration. The first phase is acceleration inside the target. In this range of intensities under the action of the ponderomotive force of the circularly polarized laser pulse electrons within the thin solid-density target can be appreciably shifted from their initial positions without strong heating, whereas the heavy ion component remains almost unchanged, thus producing a substantial charge separation field. This quasistatic field within the target may be used for effective acceleration of protons initially placed at the front side of the target, which is essentially in opposite to the TNSA regime, where protons are accelerated from the rear side. We will refer to this process as a capacitor-like acceleration phase.

The second acceleration phase sets in for thin targets when the electron layer is ponderomotively compressed for high intensity lasers down to the skin-layer thickness. In this case the effect of relativistically induced slab transparency (RIST) takes place. The crucial point of this effect is sharp dependence of transparency coefficient on the incident intensity. Thus if the incident laser pulse is short enough it results in the transmitted laser pulse with sharp intensity distribution which is able to push electrons forward as well. These electrons are capable of being emitted at distances of several microns to produce an additional charge separation field that can accelerate protons that passed through the created capacitor-like structure to higher energies.

The key point is also that the pulse that has passed through the layer may be used again in the next foil by producing potential difference in the latter and also making it transparent. The protons accelerated in the first layer get into the accelerating potential produced in the second layer and gain additional energy. And this will be refered to as the third phase of acceleration. Thus it is possible to realize the idea of cascadable proton beam acceleration. By adjusting the distance between the targets it is possible to control acceleration efficiency and proton beam properties, which is especially important in the context of practical applications. We will also demonstrate that with optimal choice of target structure the proposed scheme allows one to obtain proton beams with energies of hundreds of MeV and a few percent energy spread in laser facilities available today.

{\em Accelerating potential at RIST regime.}{\textemdash}Let us consider irradiation of a thin foil with ions heavy enough to be supposed as immobile by a superintense laser pulse. Further we will consider laser pulse to be circularly polarized as to be sure that electron dynamics is dominated by radiation pressure and also to avoid effective electron heating by ${\bf j}\times{\bf B}$ mechanism. We will also suppose the problem to be one-dimensional assuming the foil thickness to be much less than the transverse size of the laser beam. In this system one of the two regimes is realizing: if the intensity does not exceed some threshold value, an electromagnetic wave is reflected almost completely, it is a reflection regime; otherwise, essential part of the incident energy flux can be transmitted through and it is a RIST regime.

\begin{figure}
\includegraphics[width=0.5\textwidth]{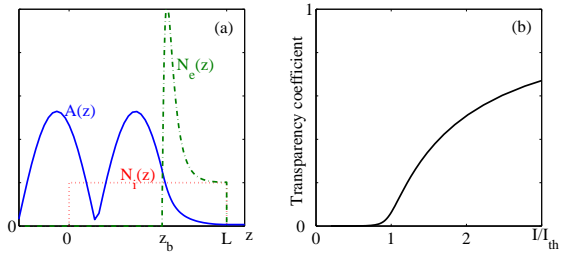}
\caption{(a) Typical stationary distribution of vector potential (blue solid curve) and of electron and ion concentrations (green dash-dotted and red dotted curves respectively). The laser energy flux is almost totally reflected, and electrons in the plasma layer are eventually pushed by the ponderomotive force that leads to formation of longitudinal electric field at the front side of the target, which itself can accelerate proton \cite{korzhimanov1}; (b) The layer transparency coefficient in dependence on incident laser intensity. The intensity is normalized to the threshold value (2)}
\end{figure}

In the reflection regime, an electromagnetic wave penetrates into the electron layer only by the thickness of skin-layer which is less than the foil width. Plasma-field distributions arose during such interaction were considered in \cite{korzhimanov2}. The analysis were based on Maxwell's equations and hydrodynamic equations for the electron component of plasma where the ion motion and plasma temperature were neglected.  An example of such a quasi-stationary structure is given in Fig. 1a. Besides, charge redistribution in the plasma results in considerable drop in electrostatic potential that may be used for acceleration of positively charged particles, in case of they pass over the layer from the front to the rear boundary. The displacement of electrons (see Fig. 1a) and, hence, the magnitude of the potential drop depends on the incident radiation intensity and increase with intensity by the law \cite{korzhimanov1}
\begin{equation}
	kz_b \approx \frac{2\sqrt{I}}{n}
\end{equation}
where $z_b$ is the coordinate of the shifted electrons boundary, $k$ is the wave number of the incident wave, $I$ is the laser intensity normalized to the so-called relativistic intensity $I_{rel}[\mbox{W/cm}^2]\approx 2.75\times 10^{18}\lambda^{-2}[\mu\mbox{m}]$, and $n$ is the unperturbed electron concentration normalized to its critical value at a given laser frequency (for overdense plasma $n>1$).

However, taking into account that $z_b$ cannot exceed the layer thickness $L$ and making use of Eq.(1) at $z_b \approx L$, we can conclude that the reflection regime occurs only when the incident intensity is less than a certain threshold value 
\begin{equation}
	I_{th} \approx \left(\frac{nkL}{2}\right)^2
\end{equation}

If the incident intensity is comparable to or exceeds the threshold value, the RIST regime is realized. The important point is that the coefficient of transparency is quite rapidly increasing with intensity (see Fig. 1b) \cite{korzhimanov2}. It means that the profile of transmitted pulse may be significantly sharpened, which results in enhanced ponderomotive force acting effectively on electrons. This force is able to push a part of electrons out of the slab thereby generating an additional accelerating field outside of the slab. The areal density of ejected electrons $\sigma$ can be estimated as follows
\begin{equation}
	\sigma = \frac{1}{4\pi c}\max\left[\frac{\partial A_{tr}(z,t)}{\partial t}\right]
\end{equation}
Here $A_{tr}(z,t)$ is the shape of transmitted pulse. Simple calculations give that transmitted pulse with intensity of the order $10^{22}$ W/cm$^2$ arising on the couple of waveperiods is able to carry away about 10 mC/cm$^2$ and produce electric field of the order of $10^{14}$ V/m which is comparable with that inside the slab. One-dimensional fully relativistic PIC code verified that the accelerating potential outside the slab may be comparable with or even more than that inside the slab. This effect occurs in the case of ultrashort ($<30$~fs) laser pulses when the transition from the reflection to the transparency is rapid.


{\em Proton acceleration in a single layer.}{\textemdash}In Fig. 2 the results of the interaction of a 30~fs (at FWHM of the Gaussian profile) laser pulse with a 100 nm gold foil are shown. The gold foil was placed at interval [0~nm; 100~nm].  The simulation used a system domain 16~$\mu$m long with absorbing boundaries for fields and particles. A laser with wavelength 1~$\mu$m and maximum intensity $10^{22}$~W/cm$^2$ enters from the left boundary and interacts with a target located at 2~$\mu$m from the left boundary. The target is composed of electrons and gold ions Au$^{6+}$ at solid density (overdense parameter $n=100$). The simulation cell size is $0.9\lambda_{D0}$, with $\lambda_{D0}=5\times10^{-8}$~cm representing the initial Debye length of the electrons. The initial electron and ion temperatures were 1~keV.

\begin{figure}
\includegraphics[width=0.45\textwidth]{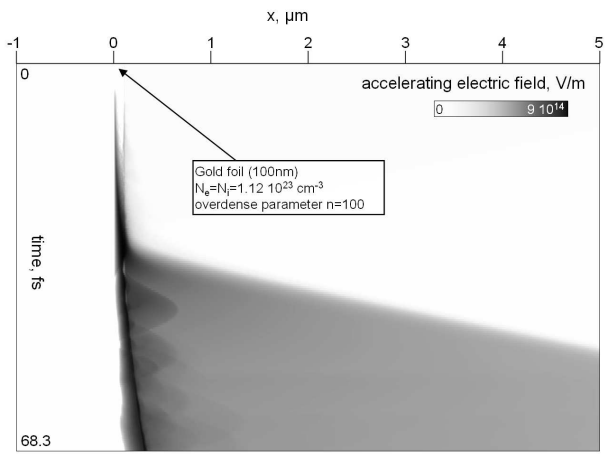}
\caption{ The space-time diagram of the longitudinal electric field dynamics in the interaction of a single gold foil and a 30~fs laser pulse of peak intensity $10^{22}$ W/cm$^2$. Gold foil initially was placed at interval [0~nm; 100~nm]. At the moment about 30~fs the transition form the refection regime to the RIST regime is clearly seen}
\end{figure}

In Fig. 2 the white-to-black shades show space-time dynamics of a longitudinal accelerating electric field where its maximum is of the order of 0.9 PV/m. Protons can gain their energy mostly in the regions where this field is large (dark regions). The protons whose trajectories pass through the dark regions, thus providing maximal final energy, have optimal initial position. As is clearly seen in the figure, two points should be emphasized. Firstly, the optimal position of the layer with accelerated protons is on the front side of the thin foil rather than on its rear one. Secondly, it is important to control the moment of injecting a proton beam in the foil because the accelerating potential we want to use is formed by the pushed out bundle of electrons only for a short time. To provide a possibility to control the moment protons are injected into a plasma layer the proton layer should be placed at some distance before the substrate of heavy ions foil.

We have observed the described dynamics in a series of numerical computations in which we found the optimal position of the proton layer. In the numerical experiment shown in Fig. 3 we used a laser pulse with the same parameters as in Fig. 2. The target comprised two layers 100 nm thick spaced apart by 200 nm. The first layer consisted of protons and electrons in equal concentration of $1.12\times 10^{22}$ cm$^{-3}$ (the overdense plasma parameter being about 10). The second layer was taken the same as in Fig.2 simulation. Our simulations demonstrated that more than half of the protons from the proton layer formed a quasi-monoenergetic proton beam that was accelerated up to energies of about 50 MeV.
\begin{figure}
\includegraphics[width=0.45\textwidth]{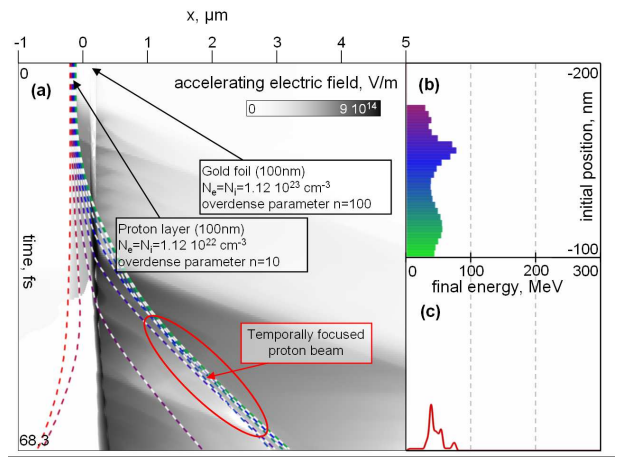}
\caption{(a) Trajectories (dashed lines) of a number of protons having different initial positions in the case of a target comprised of proton and gold layers spaced apart and irradiated by a 30 fs laser pulse of peak intensity $10^{22}$ W/cm$^2$. The black-to-white shades show a longitudinal accelerating electric field as well; (b) the diagram of final energy in dependence on initial proton position; (c) the final proton energy distribution}
\end{figure}

We would like to note that the plasma slab works also as a proton lens and we obtained the effect of temporal proton beam focusing. The focusing occurs because the faster protons in the leading part of the proton beam have smaller acceleration than the lagging protons that move with smaller velocity and get into a high accelerating field. This results in formation of a proton beam with small energy spread and highly localized in space (or in time - short proton pulse) as well.

{\em Multi-cascade regime.}{\textemdash}As was stated above one of the essential advantages of the proposed scheme is that after RIST effect occurs the laser radiation begins to penetrate through gold foil almost without any reflection. So that the process of proton acceleration may be repeated again in the next plasma layer. For this we should place the heavy ion foil following the first substrate at such a distance that the protons could get into it at the instant the laser pulse passed through the first accelerating layer produces the largest potential drop in the second layer. To achieve the best result we can fit both the interval between the layers and layer thickness. However, numerical modeling shows that quite a good result may be obtained using a second plasma layer of the same thickness as the first one. For optimal parameters at first layer the accelerated proton beam escapes it almost at the instant it becomes transparent. It takes the transmitted laser pulse propagating with the speed of light less time to pass the interval between the layers than needed for the proton beam. As some part of the incident intensity reflects from the first layer in the RIST regime it needs some time for the transmitted intensity to reach the threshold value on the second layer. This time delay allows one to synchronize the moment of proton beam arrival to the second slab with the moment when the RIST effect occurs in this slab. This means that there exists a distance between the layers at which the proton beam transits into the second layer at the instant RIST effect occurs. It is also important to note that accelerating layers do not work independently. The dynamics of laser-slab interaction is affected by electrons expelled out of the previous layers. So this dynamics may be quite a complex in nature.

After the optimization of single-layer acceleration process the positions of other layers may be optimized one after another in order of their interaction with laser pulse. We used this method in our numerical experiment to construct a target containing four layers. In the constructed model, the proton beam consisted of $10^9$ particles per $\rm\mu m^2$ is formed in the first accelerating layer, passes all the four cascades, is successively accelerated in them, and as a result has the output energy of about 220 MeV with rather small energy spread of just 2.3\% (see Fig.4c).

The results of numerical experiment on generation of accelerated proton beam on a target comprising four accelerating layers are demonstrated in Fig.4. Like before, we considered a laser pulse with Gaussian profile, peak intensity $10^{22}$ W/cm$^2$, and 30 fs duration. The target consisted of a proton containing layer and four identical accelerating layers. The intervals between the targets are shown in Fig.4.

\begin{figure}
\includegraphics[width=0.45\textwidth]{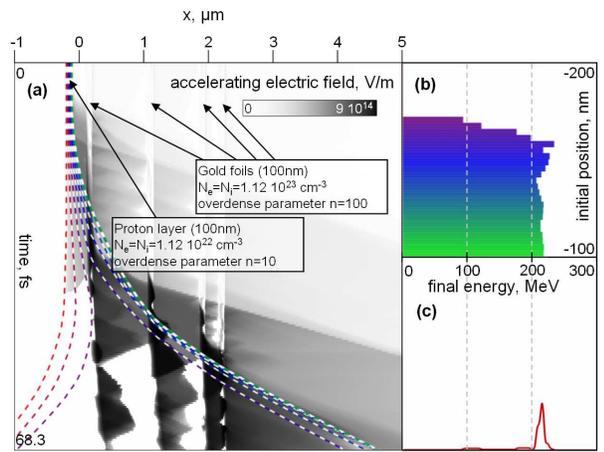}
\caption{Proton acceleration in the case of target consisted of four gold foils. Panels (a-c) show the same as in Fig.3}
\end{figure}

Fig. 4a depicts trajectories of different protons and a charge separation accelerating field in the plane of longitudinal coordinate and time. Of course, numerical fitting of parameters cannot determine accurately parameters of a real target but it gives a good illustration of the capabilities of the proposed scheme of proton acceleration. In other words, the carried out qualitative numerical experiments enable us to state that multi-cascade targets with optimally chosen layer thicknesses and intervals between them may be used in real experiments for producing mono-energetic proton beams with energies up to several hundreds of MeV.

{\em Conclusion.}{\textemdash}We have examined a new laser driven proton or light ion acceleration mechanism that results from the interaction of high intensity circularly polarized laser pulses with thin foils. This mechanism relies on the acceleration by longitudinal charge separation fields generated in the ponderomotively predominant regime of laser-plasma interaction, which proceeds through two distinct phases. In the first phase, the longitudinal accelerating field is produced inside the target due to electron displacements by the ponderomotive force of the incident laser pulse. When the electron displacement reaches the rear side the thin foil is rapidly becoming transparent allowing the leading edge of the laser pulse passing through to be very sharp. This second phase, which we call as RIST effect, results in the enhanced ponderomotive force acting on the rear side electrons expelling them out of the target. These expelling electrons can contribute significantly to the accelerating field. 

The key point of the RIST regime is that the slab becomes almost transparent for radiation and thus allows it to pass through. So that this radiation may be used at the next slab and so forth. This effect underlies the proposed regime of multi-cascade proton acceleration. The number of possible cascades is determined by laser pulse properties. Firstly, it energy should be enough to go through all cascades as it loses some part of energy on each plasma slab. Secondly, taking into account that the laser pulse has maximal intensity only in focal spot the total thickness of the target should not exceed the Rayleigh length.

Particle-in-cell simulations show that this regime of ion acceleration is especially effective in the intensity range of $10^{21}-10^{22}$~W/cm$^2$ and pulses with ultrashort durations, less than 30~fs, and allows to generate actually a monoenergetic beams of ions. The main advantages of this mechanism of acceleration is that the ion energy can be eventually multiplied by using several space-separated foils. In simulation presented above the proton beam with energy of 220 MeV and energy spread of 3\% was generated in a multi-cascade target comprising of four identical gold foils of 100 nm thickness.

\end{document}